\documentclass[12pt,onecolumn,amsmath,amssymb]{revtex4}

\begin{document}

\title{Sufficient integral criteria for instability
of the free charged surface of an ideal liquid}

\author{Nikolay M. Zubarev}
\email{nick@ami.uran.ru}
\author{Olga V. Zubareva}

\affiliation{Institute of Electrophysics, Ural Branch, Russian
Academy of Sciences,\\ 106 Amundsen Street, 620016 Ekaterinburg, Russia}

\begin{abstract}

Applying the method of integral estimates to the analysis of three-wave
processes we derive the sufficient criteria for the hard loss of
stability of the charged plane surface of liquids with different physical
properties. The influence of higher-order wave interactions on the
instability dynamics is also discussed. 

\end{abstract}

\maketitle

As we know \cite{1,2,3,4}, the dispersion relation for electrocapillary waves on a
charged liquid surface has the following form:
\begin{equation}
\omega^2=gk+\frac{\alpha}{\rho}\,k^3-\frac{P}{4\pi\rho}\,k^2
\end{equation}
where $k=|\mbox{\bf k}|$ is the wave number, $\omega$ is the frequency,
$g$ is the acceleration of gravity, $\alpha$ is the surface tension,
$\rho$ is the mass density, and $P$ is an external control parameter,
depending on liquid nature. So, $P=E^2$ for an ideal perfectly conducting
liquid (liquid metal in applications), where $E$ is the external electric
field strength. For an ideal dielectric liquid the parameter $P$ equals
the expression $E^2(\varepsilon-1)^2/(\varepsilon^2+\varepsilon)$, where
$\varepsilon$ is the permittivity. For liquid helium and liquid
hydrogen it holds $P={E_+}^2+{E_-}^2$, where $E_+$ and $E_-$ designate the
electric field strength above and below the fluid surface, respectively.
One can see from Eq.~(1) that if the control parameter $P$ exceeds the
critical value,
$$
P_c=8\pi\sqrt{g\alpha\rho},
$$
then $\omega^2<0$ and, as a consequence, an aperiodic instability
develops. Thus, the condition $P>P_c$ is the criterion for the surface
instability with respect to infinitesimal perturbations of the surface
shape and the velocity field.

As shown in Refs.~\cite{5,6,7}, nonlinear interactions between three standing
waves, which form a hexagonal structure, can lead to the hard loss of
stability of a charged liquid surface. Then, even if the value of the
control parameter is subcritical, i.e., $P<P_c$, a fairly large-amplitude
perturbation can remove the system from equilibrium, resulting either in
the formation of a perturbed stationary surface profile or in infinite 
growth of the amplitudes of surface perturbations in accordance with
character of higher-order wave processes. An important task of physical
significance is therefore to obtain criteria for instability of the plane
surface with respect to perturbations with finite amplitudes, namely,
to determine the initial conditions (surface configuration and
distribution of the velocity field) which lead to the development of
an instability.

In the present Letter we construct and discuss such criteria for the case
of a conducting liquid in an external electric field (all our
results may be extended to other fluids) by using the method of
majoring equations, which was applied previously to the nonlinear
Schr$\ddot{\mbox{o}}$dinger equation (see, for example, Refs.~\cite{8,9}), the
nonlinear Klein-Gordon equation \cite{10,11}, different modifications of the
Boussinesq equation \cite{12}, and so on. This method allows us to derive a
number of sufficient integral criteria for the loss of stability of the
plane charged liquid surface, with most of them corresponding to the
subcritical values of the control parameter $P$, when the surface is
stable in the linear approximation and the instability onset is caused by
nonlinear processes.

So, let us consider the irrotational motion of a perfectly conducting
ideal liquid of infinite depth with a free surface $z=\eta(x,y,t)$
in an external electric field $E$ directed along the $z$ axis.
The velocity potential $\Phi$ and the potential of the electric field
$\varphi$ obey the Laplace equations
$$
\nabla^2\Phi=0, \qquad \nabla^2\varphi=0
$$
with the conditions at infinity:
$$
\Phi\to 0, \qquad z\to-\infty,
$$
$$
\varphi\to -Ez, \qquad z\to\infty.
$$
The conditions on the free surface are
$$
\frac{\partial\Phi}{\partial t}+\frac{(\nabla\Phi)^2}{2}=
\frac{(\nabla\varphi)^2-E^2}{8\pi\rho}+
\frac{\alpha}{\rho}\,\nabla_{\!\!\bot}\cdot
\frac{\nabla_{\!\!\bot}\eta}{\sqrt{1+(\nabla_{\!\!\bot}\eta)^2}}-
g\eta, \qquad z=\eta,
$$
$$
\frac{\partial\eta}{\partial t}=\frac{\partial\Phi}{\partial z}
-\nabla_{\!\!\bot}\eta\cdot\nabla_{\!\!\bot}\Phi,
\qquad z=\eta,
$$
and, since the surface of a conducting liquid is equipotential,
$$
\varphi=0, \qquad z=\eta.
$$

The functions $\eta(x,y,t)$ and $\phi(x,y,t)=\Phi|_{z=\eta}$ are
canonically conjugate \cite{13}, so that the equations of motion take the
Hamiltonian form,
$$
\frac{\partial\psi}{\partial t}=-\frac{\delta H}{\delta\eta},
\qquad
\frac{\partial\eta}{\partial t}=\frac{\delta H}{\delta\psi},
$$
where the Hamiltonian coincides with the total energy of the system:
\begin{equation}
H=H_{\mbox{\normalsize kin}}+H_{\mbox{\normalsize pot}},
\end{equation}
\begin{equation}
H_{\mbox{\normalsize kin}}=\int\limits_{z\leq\eta}\frac{(\nabla\Phi)^2}{2} d^3 r,
\end{equation}
\begin{equation}
H_{\mbox{\normalsize pot}}=-\int\limits_{z\geq\eta}\frac{(\nabla\varphi)^2}{8\pi\rho} d^3 r+
\int\left[\frac{g\eta^2}{2}+
\frac{\alpha}{\rho}\left(\sqrt{1+(\nabla_{\!\!\bot}\eta)^2}-1\right)\right]
d^2 r.
\end{equation}
It is possible to express $H$ explicitly in terms of the canonical
variables. Rewriting the Hamiltonian in the form of a surface integral
with the help of Green's formulas and expanding the integrand in powers
series of $\psi$ and $\eta$, we obtain
\begin{equation}
H_{\mbox{\normalsize kin}}=
\int\frac{\psi}{2}\left(\hat T_+\hat k\hat T_+^{-1}\psi-
\nabla_{\!\!\bot}\eta\cdot\hat T_+\nabla_{\!\!\bot}\hat T_+^{-1}\psi\right) d^2 r,
\end{equation}
$$
H_{\mbox{\normalsize pot}}=
-\int\frac{E^2\eta}{8\pi\rho}\left(\hat T_-\hat k\hat T_-^{-1}\eta+
\nabla_{\!\!\bot}\eta\cdot\hat T_-\nabla_{\!\!\bot}\hat T_-^{-1}\eta\right) d^2 r
$$
\begin{equation}
+\int\left[\frac{g\eta^2}{2}+
\frac{\alpha}{\rho}\left(\sqrt{1+(\nabla_{\!\!\bot}\eta)^2}-1\right)\right]
d^2 r.
\end{equation}
Here $\hat k$ is the two-dimensional integral operator with a difference
kernel whose Fourier transform is equal to the absolute value of the wave
vector ($\hat{k}e^{i\mbox{\small\bf kr}}=|\mbox{\bf k}|e^{i\mbox{\small\bf
kr}}$), and the nonlinear operators $\hat T_\pm$ are defined by the
expressions
$$
\hat T_\pm=\sum_{n=0}^{\infty}\frac{(\pm\eta)^n\hat k^n}{n!}.
$$

For values of the control parameter $\epsilon=(P-P_c)/P_c$ close to
threshold ($|\epsilon|\ll 1$), as is clear from the dispersion relation
(1), the surface perturbations with $k\approx k_0=\sqrt{g\rho/\alpha}$ 
have the maximum linear growth rates. Then the nonlinear dynamics of the
surface perturbations can be effectively studied with the help of the 
amplitude equation approach. Assuming the surface-slope angles to be
small, $|\nabla_{\!\!\bot}\eta|$, we introduce the slowly varying
amplitudes $A_j$, $j=1,2,3$, by means of the substitutions
$$
\eta(\mbox{\bf r},t)=\frac{2}{3k_0}\sum_{j=1}^3
A_j (x_j,y_j,t)\exp(i\mbox{\bf k}_j\mbox{\bf r})+\mbox{c.c.},
$$
$$
\psi(\mbox{\bf r},t)=\frac{2}{3k_0^2}\sum_{j=1}^3
\frac{\partial A_j(x_j,y_j,t)}{\partial t}
\exp(i\mbox{\bf k}_j\mbox{\bf r})+\mbox{c.c.},
$$
where the wave vectors $\mbox{\bf{k}}_j$ with $|\mbox{\bf{k}}_j|=k_0$ make
angles of $2\pi/3$ with each other, and the variables $x_j$, $y_j$ form
the orthogonal coordinate systems with $x_j$ axis directed along the wave
vectors $\mbox{\bf{k}}_j$. This representation for $\eta$ and $\psi$
corresponds to a hexagonal structure of the surface perturbations, which
is preferred at the initial stages in the development of the instability. 

Substituting the relations for $\eta$ and $\psi$ into Eqs.~(2), (5) and
(6), in the leading order we get the following expression for the averaged
Hamiltonian: 
\begin{equation}
H=\int\left[\sum_{j=1}^3\left(|{A_j}|_t^2+
|\hat L_j A_j|^2-
\epsilon |A_j|^2\right)-A_1A_2A_3-A_1^*A_2^*A_3^*\right] d^2 r,
\end{equation}
where we convert to dimensionless variables,
$$
\mbox{\bf r}\to \mbox{\bf r}/(\sqrt{2}k_0),
\qquad t\to t/\sqrt{2gk_0},\qquad
H\to 8gH/(9k_0^2),
$$
and introduce the linear differential operators
$$
\hat L_j = \frac{\partial}{\partial x_j}-
\frac{i}{2}\frac{{\partial}^2}{\partial y_j^2}, \qquad j=1,2,3.
$$
The equations for the complex amplitudes $A_j$ corresponding to this
Hamiltonian are
\begin{equation}
\frac{{\partial}^2A_j}{\partial t^2}=\epsilon A_j+
{\hat L_j}^2 A_j+
\frac{A_1^*A_2^*A_3^*}{A_j^*}, \qquad j=1,2,3.
\end{equation}
It should be noted that the multiplier
$p=(\varepsilon-1)/(\varepsilon+1)$ and the multiplier
$s=({E_+}^2-{E_-}^2)/P_c$ appear before the nonlinear terms in the
right-hand sides of Eqs.~(8) for dielectric liquid and for liquid helium,
respectively (the equations reduce to the form (8) by scaling the
amplitudes $A_j\to A_j/p$ or $A_j\to A_j/s$).

Let us find the sufficient criteria for unlimited growth of the
amplitudes $A_j$ in a finite time, i.e., the criteria of the blow-up for
Eqs.~(8), which describe the nonlinear interactions between three standing
waves. Consider the time evolution of the following positive quantity:
$$
X=\sum_{j=1}^3 X_j, \qquad X_j(t)=\int|A_j|^2 d^2r.
$$
Differentiating $X$ twice with respect to $t$ and then making use of
Eqs.~(8), we get
$$
X_{tt}=\int\left[2\sum_{j=1}^3\left(|{A_j}|_t^2-
|\hat L_j A_j|^2
+\epsilon|A_j|^2\right)+3A_1A_2A_3+3A_1^*A_2^*A_3^*\right] d^2 r.
$$
Excluding the cubic terms from the integrand with the help of the
expression (7), we come to the relation
\begin{equation}
X_{tt}+3H=-\epsilon X+
\sum_{j=1}^3\int\left[5{|A_j|}_t^2+
|\hat L_j A_j|^2 \right] d^2r.
\end{equation}
It follows from the integral H$\ddot{\mbox{o}}$lder inequality for the
functions $|A_j|$ and $|{A_j}|_t$ that
$$
4X_j\int{|A_j|}_t^2 d^2r\geq {X_j}_t^2.
$$
On the other hand, the algebraic Cauchy inequality yields
$$
\left(\sum_{j=1}^3 X_j\right)\cdot
\left(\sum_{j=1}^3{{X_j}_t}^2/X_j\right)\geq
\left(\sum_{j=1}^3 {X_j}_t\right)^2.
$$
As a consequence, we can estimate the second term in the right-hand side of
Eq.~(9):
$$
\sum_{j=1}^3\int|A_j|_t^2 d^2 r\geq\frac{{X_t}^2}{4X}.
$$
In addition, taking into account that
$$
\int|\hat L_j A_j|^2 d^2r\geq 0,
$$
we obtain the following second order differential inequality:
\begin{equation}
X_{tt}+3H\geq \frac{5}{4}\frac{{X_t}^2}{X}-\epsilon X.
\end{equation}
Solving the inequality one may find the sufficient conditions under which
the integral value $X$ becomes infinite in a finite time. It should be
noted that the similar majoring inequalities were derived in
Refs.~\cite{8,9,10,11,12} 
as a result of investigation of the blow-up in different well-known
nonlinear partial differential equations.

Under the substitution, $Y=X^{-1/4}$, inequality (10) takes the
Newtonian form
\begin{equation}
Y_{tt}\leq-\frac{\partial P(Y)}{\partial Y}, \qquad
P(Y)=-\frac{1}{8}\left(\epsilon Y^2+H Y^6\right),
\end{equation}
where $Y$ can be considered as a coordinate of some particle and the
function $P(Y)$ plays the role of the potential. Let the particle velocity
$Y_t$ be negative (in this case $X_t>0$). Then, multiplying the inequality
(11) by $Y_t$ and integrating it over time, we get
$$
U_t(t)\geq 0, \qquad U(t)={Y_t}^2/2+P(Y),
$$
i.e., the particle energy $U$ increases with time. It is clear that if the
condition $U_t=0$ holds, which corresponds to the equality sign in the
expression (11), and the particle does not encounter a potential barrier,
then it reaches the origin, i.e., the point $Y=0$, and, consequently,
the positive-definite quantity $X$ becomes infinite. The collapse takes
place
\begin{itemize}
\item[(a)]
if $\epsilon<0$ and $H>0$, provided that
$Y(t_0)<|\epsilon|^{\frac{1}{4}}/(3)^{\frac{1}{4}}$ and
$12U(t_0)\leq|\epsilon|^{\frac{3}{2}}/(3H)^{\frac{1}{2}}$;
\item[(b)]
if $\epsilon<0$ and $H>0$, provided that
$12U(t_0)>|\epsilon|^{\frac{3}{2}}/(3H)^{\frac{1}{2}}$;
\item[(c)]
if $\epsilon<0$ and $H\leq 0$;
\item[(d)]
if $\epsilon\geq 0$, provided that $U(t_0)>0$,
\end{itemize}
where $t=t_0$ corresponds to the initial moment. The collapse time $t_c$
at which the amplitudes $A_j$ go to infinity can be estimated from above
as follows,
$$
t_c\leq t_0+\!\!\int\limits_0^{Y(t_0)}\!\!\frac{dY}
{\sqrt{2U(t_0)-2P(Y)}}.
$$
Note that the condition $Y_t(0)<0$ is optional for the cases (a) and (c).
Since $U_t(t)\leq 0$ for $Y_t>0$, in these two cases the particle always
reaches the point $Y=0$ after the reflection from the potential barrier.

It is important that the conditions (a)--(d) may serve as the sufficient
criteria for the instability of the plane surface of a
conducting liquid in the near-critical electric field with respect to
finite amplitude perturbations. This distinguishes our criteria from the
simplest criterion for linear instability, $P>P_c$, which corresponds
to infinitesimal perturbations. Actually, the conditions (a), (b) and (c)
are relative to the case of the subcritical electric field strength,
$P<P_c$, when the surface is stable in the linear approximation. It is
obvious that we deal with hard excitation of the electohydrodinamic
instability.

Thus, we have shown that if the conditions (a)--(d) are valid, then
the equations (8) describe the unlimited growth of the amplitudes $A_j$.
However, the applicability of Eqs.~(8), i.e., the possibility of limiting
the treatment to three-wave processes, assumes that the perturbation
amplitudes are small ($|A_j|$ are the order of magnitude of the control
parameter $\epsilon$). The question arises as to whether the neglected
higher-order nonlinearities stabilize the instability, or, on the
contrary, higher-order wave processes promote an explosive growth of the
amplitudes. Notice that both the experimental data \cite{14} and the results of
the numerical calculations \cite{15} indicate that the higher-order
nonlinearities have the destabilizing influence. 

Let us show that, in particular, the four-wave interaction does not
saturate the explosive instability. Consider the simplest case when the
perturbation amplitudes do not depend on the spatial variables $x$ and
$y$, and the surface configuration is given by
$$
\eta(\mbox{\bf r},t)=\sum_{j=1}^3
(\xi_j+2 k_0 \xi^2_j)+\frac{k_0}{2}(5\sqrt{3}+6)
(\xi_1 \xi_2^*+\xi_2 \xi_3^*+\xi_3 \xi_1^*)+\mbox{c.c.},
$$
$$
\xi_j(\mbox{\bf r},t)=a_j (t)\exp(i\mbox{\bf k}_j\mbox{\bf r}),
$$
where the nonlinear interactions between fundamental and combination
harmonics, $k_0\leftrightarrow 2k_0$ and $k_0\leftrightarrow \sqrt{3}k_0$,
are taken into account. Substituting these expressions into (6), 
we obtain the fourth-order correction for the potential energy (4):
$$
H_{\mbox{\normalsize pot}}^{(4)}=
-gk_0^2\sum_{j=1}^3\int \left[\frac{11}{4}|a_j|^4 +
\frac{1}{2}(25\sqrt{3}+13)|a_j|^{-2}\prod_{j=1}^3|a_j|^2\right] d^2r.
$$
This functional is the negative-definite quantity,
$H_{\mbox{\normalsize pot}}^{(4)}\leq 0$. Consequently, we can assume
that the potential energy $H_{\mbox{\normalsize pot}}$ decreases
indefinitely as the perturbations grow. Then the kinetic energy
$H_{\mbox{\normalsize kin}}$, which is the positive-definite quantity,
increases infinitely (see Eqs.~(2) and (3)). Thus, the higher-order
nonlinearities do not retard the explosive growth of the amplitudes in the
model (8), and the integral criteria (a)--(d) can be considered as the
sufficient criteria for the unlimited growth of the perturbations of a
conducting liquid surface in an applied electric field.

As for a dielectric liquid in the near-critical electric field, we have
$$
H_{\mbox{\normalsize pot}}^{(4)}=
-gk_0^2\sum_{j=1}^3\int \left[\left(4p^2-\frac{5}{4}\right)|a_j|^4 +
\frac{1}{2}\left(15-12\sqrt{3}+p^2(37\sqrt{3}-2)\right)
|a_j|^{-2}\prod_{j=1}^3|a_j|^2\right] d^2r.
$$
It can readily be seen that $H_{\mbox{\normalsize pot}}^{(4)}\leq 0$ for
arbitrary $a_j$ if $\varepsilon\geq\varepsilon_{2D}$ holds, where
$\varepsilon_{2D}\approx3.53$ (at this value of the permittivity
the hard instability regime changes to the soft one in 2D geometry \cite{16}).
Then the relations (a)--(d) represent the criteria for the blow-up-type
dynamics of the surface perturbations. If 
$1<\varepsilon\leq\varepsilon_{3D}$ holds, where
$\varepsilon_{3D}\approx2.05$ (the functional 
$H_{\mbox{\normalsize pot}}^{(4)}$ with $|a_1|=|a_2|=|a_3|$ changes
the sign precisely at this value of the permittivity $\varepsilon$ \cite{6}),
then $H_{\mbox{\normalsize pot}}^{(4)}\geq 0$ and, consequently, the
four-wave processes can stabilize the instability, resulting in the
appearance of the stationary hexagonal structures. In this case the
conditions (a)--(c) are the criteria for the hard excitation of the
stationary wave patterns on the free surface of an ideal dielectric
liquid. 

For liquid helium (or hydrogen) with a charged surface it holds
$$
H_{\mbox{\normalsize pot}}^{(4)}=
-gk_0^2\sum_{j=1}^3\int \left[\left(4s^2-\frac{5}{4}\right)|a_j|^4 +
\frac{1}{2}\left(16\sqrt{3}-29+s^2(9\sqrt{3}+42)\right)
|a_j|^{-2}\prod_{j=1}^3|a_j|^2\right] d^2r.
$$
One can find that $H_{\mbox{\normalsize pot}}^{(4)}\leq 0$ for
$|s|\geq |s_{2D}|$, where $s_{2D}\approx 0.56$ (at this value of the
parameter $s$ the functional $H_{\mbox{\normalsize pot}}^{(4)}$ changes
its sing in 2D case \cite{4}), and $H_{\mbox{\normalsize pot}}^{(4)}\geq 0$ for
$|s|\leq|s_{3D}|$, where $s_{3D}\approx 0.33$.

\medskip
The authors are grateful to E.A.~Kuznetsov for stimulating discussions,
and also to A.M.~Iskoldsky and N.B.~Volkov for their interest in this
work. The work was supported by the Russian Fund for Fundamental Research
(Project No.~00-02-17428) and, partly, by the INTAS Fund (Project
No.~99-1068).

\end{document}